\documentclass[12pt,twoside]{article}
\usepackage[numbers,sort&compress]{natbib}
\usepackage{pbox}

\newcommand{\absatz}{\vspace*{2ex}\noindent}

\begin{document}

\textbf{-- To Whom It May Concern --} \hfill Trento, 1 August 2013

\absatz \absatz \textbf{Concerning: \pbox[t]{\linewidth}{Theory Viewpoint on
    Extracting Nucleon Polarisabilities\\[0.5ex] in Low-Energy Compton
    Scattering}}

\absatz

During the workshop \textsc{Compton Scattering off Protons and Light Nuclei:
  Pinning Down the Nucleon Polarisabilities} at the ECT* (Trento, Italy), we
have been asked by our experimental colleagues to summarise the present common
theoretical understanding on the feasibility of extracting the static dipole
scalar and spin polarisabilities from low-energy Compton scattering off the
proton and light nuclei. These quantities parametrise the deformation of the
nucleon in external electric and magnetic fields, and lattice QCD results for
them are emerging. Besides being fundamental properties of the nucleon, they
play an important role in the Lamb shift of muonic Hydrogen as well as in
radiative corrections to the proton charge radius, and provide the biggest
source of uncertainty in theoretical determinations of the proton-neutron mass
shift. Spin polarisabilities parametrise the optical activity of the nucleon
and test its spin degrees of freedom. Scattering on light nuclei allows one to
differentiate between proton and neutron values, and thus to study chiral
symmetry breaking.

As highlighted in the Long-Range Plans in the USA (NSAC 2007, NAS 2012) and
Europe (NuPECC 2010), this vibrant and renewed theoretical interest prompted a
new generation of high-accuracy facilities with unpolarised and polarised
photon beams and targets to focus on Compton scattering. Interpreting such
data needs commensurate theoretical support for interpretations with minimal
theoretical bias.

Our credentials are publications in a range of theoretical approaches to the
problem, namely several variants of both Dispersion Relations and Effective
Field Theories~\cite{Alarcon:2013cba, Babusci:1998ww, Beane:2004ra,
  Beane:1999uq, Birse:2012eb, Choudhury:2007bh, Drechsel:2002ar,
  Griesshammer:2013vga, Griesshammer:2012we, Krupina:2013dya, Lensky:2014efa,
  Lensky:2009uv, L'vov:1996xd, McGovern:2012ew, Pascalutsa:2002pi, Pasquini:2010zr, Pasquini:2007hf, Wolf:2001}. We agree on the following
statements.

Compton scattering up to the first resonance region can roughly be divided
into three regimes of different theoretical interest. The transition from one
regime to another is of course gradual rather than sudden. 

In the first regime, comfortably below the single pion production threshold,
our theoretical approaches contain very similar physics. Therefore, an
extraction of static polarisabilities by running cross sections and other
observables down to zero energy suffers only from minimal discrepancies
between the different theoretical approaches. At these scales, this running is
dominated by the physics of the pion cloud, which is for these energies
adequately captured by each approach. We therefore anticipate that when the
same data is used by different approaches, their values for the static
polarisabilities will agree very well. Scalar polarisabilities should be
extractable with high theoretical accuracy and minimal theory error. The same
holds for the spin-polarisabilities -- if the necessary experimental accuracy
can be reached. At present, single and double polarised data is sorely missed.

In the second regime, around and above the pion production threshold, the
sensitivity to the spin polarisabilities is increased. The different
theoretical approaches still largely agree, but different physics at this
scale leads to some discrepancies. Data in this regime will help to understand
and resolve these issues and provide first values for the spin
polarisabilities, triggering even more theoretical efforts.

In the third regime, around and above the $\Delta(1232)$ resonance, all
theoretical approaches gradually become less reliable for different reasons.
In Dispersion Relations, an accurate inclusion of the two-pion production
process in present formulations becomes crucial and is subject to further
investigation. In Effective Field Theories, the dimensionless expansion
parameter starts to approach unity, indicating increasingly worse convergence.
At present, all theoretical approaches must thus resort to well-motivated but
not fully controlled approximations. Concurrently, sensitivity to the static
polarisabilities decreases substantially. Taken together, this makes their
extraction from data at these energies less reliable. Instead, one gains
access to details of $\Delta(1232)$ resonance properties, as well as potential
information on the degrees of freedom exchanged between photons and the
nucleon in the t-channel.
Data in this regime will help improve our theoretical understanding of the
lowest nucleonic resonance, of excitations with the same quantum numbers as
the QCD vacuum, and of the interplay between the two.

In summary, we strongly support our experimental colleagues in their goal to
provide data of great relevance and high accuracy with reliable systematic
uncertainties. Only a concerted effort of both experiment and theory will
improve our understanding of the two-photon response of the nucleon. We thus
look forward to study with them the sensitivities of both unpolarised and
polarised cross sections and asymmetries of protons and light nuclei on
scalar and spin polarisabilities.  This will lead to strong experimental
proposals which address these fundamental questions. In the longer term, we
welcome a complete set of experiments up to the pion production threshold to
disentangle detailed information from the energy dependence of the Compton
multipoles.

\vspace*{4ex}

Harald W.~Grie{\ss}hammer (George Washington University, USA)
\\
Anatoly I.~L'vov (Lebedev Physical Institute, Russia)
\\
Judith A.~McGovern (University of Manchester, UK)
\\
Vladimir Pascalutsa (University of Mainz, Germany)
\\
Barbara Pasquini (University of Pavia, Italy)
\\
Daniel R.~Phillips (Ohio University, USA)

\end{document}